\documentclass[review]{elsarticle}
\pdfoutput=1
\usepackage{lineno,hyperref}
\usepackage{graphicx}
\usepackage[linesnumbered,ruled,lined]{algorithm2e}
\usepackage{geometry}
\usepackage{float}
\usepackage{tikz}
\usepackage{epstopdf}
\usepackage{amssymb}
\usepackage{amsmath}
\usepackage{subcaption}
\usepackage{footnote}
\usepackage{array}
\usepackage{cases}
\usepackage{multirow}
\usepackage{color}
\usepackage{hyperref}
\graphicspath{ {./Figure/} }
{\color{red}\journal{Elsevier}}
\begin{document}
%
%
\begin{frontmatter}
		\title{Liquid sloshing behaviours in an elastic tank and suppression effect of baffles}
		\author[myfirstaddress]{Chenxi Zhao}
		\ead{chenxi.zhao@tum.de}
            \author[myfirstaddress]{Yan Wu}
		\ead{derekwu1026@163.com}
		\author[myfirstaddress]{Yongchuan Yu}
		\ead{yongchuan.yu@tum.de}
		\author[myfirstaddress]{Oskar J. Haidn}
		\ead{oskar.haidn@tum.de}
		\author[mysecondaryaddress]{Xiangyu Hu \corref{mycorrespondingauthor}}
		\ead{xiangyu.hu@tum.de}
		\address[myfirstaddress]{Chair of Space Propulsion and Mobility, 
			Technical University of Munich, 85521 Ottobrunn, Germany}
		\address[mysecondaryaddress]{Chair of Aerodynamics and Fluid Mechanics, 
			Technical University of Munich, 85748 Garching, Germany}
		\cortext[mycorrespondingauthor]{Corresponding author. }

\begin{abstract}
In this paper, a fluid-structure interaction (FSI) framework based on the smoothed particle hydrodynamics (SPH) method is 
employed to investigate the forces and deformations experienced by LNG tanks during liquid sloshing. As a 
Lagrangian approach, the SPH method offers the advantage of accurately modelling free-surface flow. The fluid phase 
consisting of water and air is modelled as a multi-phase system for getting closer to real transport situations.
Additionally, the application of FSI within a single framework reduces data transfer discrepancies between 
fluid dynamics and solid mechanics.  To validate the reliability of the numerical methodology, the simulation results 
about the free surface elevation and wave profiles are compared with experimental data. Subsequently, ring baffles and vertical 
baffles are introduced separately. While the degree of force acting on the tanks is assessed, the anti-sloshing 
effectiveness of baffles on sloshing suppression and the variations in stress and strain distributions are 
evaluated. Further, to compare the influence of the material properties of baffles on sloshing phenomena, 
the rigid baffle and elastic baffle with different Young's moduli are immersed in the liquid. The results 
indicate that in this LNG tank configuration, the closer the baffle properties align with rigidity, 
the more effective the sloshing inhibition. 
\end{abstract}

\begin{keyword}
Tank sloshing \sep SPH \sep FSI \sep deformable structure \sep multi-phase flow
\end{keyword}

\end{frontmatter}
%
%
\section{Introduction} \label{section: introduction}

Liquefied natural gas (LNG), as a relatively cleaner energy source, finds wide applications in urban gas 
supply, industrial manufacturing and power generation. The marine transportation of LNG has always been a 
topic of paramount importance. Due to the complicated and changeable conditions at the seas, the sloshing 
behaviours of the liquid in LNG tanks with the effects of wind and waves deserve attention. 
Several experiments have been conducted to replicate sloshing phenomena in horizontal tanks under external 
excitation \cite{grotle2017experimental, liu2018numerical}, and the application of baffles was demonstrated 
to exert effective influences on mitigating the liquid 
sloshing \cite{goudarzi2010investigation, xue2013effects, jin2014experimental, nasar2019sloshing}. 

On the other hand, the experimental work is usually constrained by its expenses in time and space, 
whereas the theoretical analyses are suitable for linear or weakly nonlinear sloshing problems 
\cite{aly2015numerical, faltinsen1974nonlinear, faltinsen1978numerical, goudarzi2010investigation}. 
The advent of numerical simulation technology has expanded the toolkit for dealing with highly nonlinear 
sloshing motions. Extensive numerical investigations have been carried out regarding the interaction between 
tank sloshing and various types of baffles. In the aspect of grid-based methods, Cho et al. 
\cite{cho2004numerical} and Biswal et al. \cite{biswal2006non} studied the rigid baffles in two-dimensional 
tanks with the finite element method (FEM), and improved the free-surface tracking technique. 
Many researchers have utilized the finite difference method 
(FDM) \cite{kim2001numerical, chen2005time, xue2011numerical, paik2014fluid, unal2019liquid} or the finite 
volume method (FVM) \cite{threepopnartkul2012effect, saripilli2018numerical, yang2021numerical} to 
solve Navier-Stokes equations of fluid dynamics in tanks. Tang et al. \cite{tang2021numerical} compared the 
effect of different height allocation schemes of vertical baffles with the $k$-$\epsilon$ model, 
which indicated that the unequal baffle height has greater potential for sloshing suppression. 
Xue et al. \cite{xue2020seiche} conducted numerical investigations on the seiche oscillations of layered 
fluids, and their findings suggested that the vertical baffles accelerate the wave decay and also affect the 
dominant response frequencies. As for meshless methods, Hwang et al. \cite{hwang2016numerical} employed a 
modified moving particle simulation approach to simulate the sloshing flows in partially filled tanks with 
elastic baffles. Aly et al. \cite{aly2015numerical} established an incompressible SPH method with the Neumann 
pressure boundary to improve the calculation accuracy of highly nonlinear sloshing problems in long-time 
simulation, and performed this method on liquid sloshing with floating and middle baffles. 
Cao et al. \cite{cao2014sloshing} adopted the SPH method and found that when the ratio of the height of the 
baffle to that of the still liquid is over 0.5, it leads to an increase in the lowest natural frequency to 
reduce liquid sloshing. Additionally, simulations have explored tanks equipped with other types of baffles, 
such as porous baffles \cite{george2020anti, nimisha2022effective}, 
T-shaped baffles \cite{unal2019liquid, zhu2021effect, trimulyono2022investigation} and 
orifice baffles \cite{xue2012numerical}.

Violent oscillations can impose significant pressure loads on tanks, leading to structural instability or even 
failure. It is therefore crucial to figure out the hydrodynamic forces exerted by the fluid onto the walls. 
Lee et al. \cite{lee2007parametric} numerically studied the sensitivity of a series of parameters to the 
impact pressure on LNG tanks to point out the compressibility of ullage spaces as an essential parameter. 
A simple formula was derived by Chen et al.\cite{chen2005time} to calculate the horizontal force coefficient 
$C_{f}$ on the tank walls, indicating that at high excitation frequencies, $C_{f}$ is dominated by 
mass effects, otherwise by the free-surface displacement. Besides, Delorme et al. \cite{delorme2009set} introduced 
an SPH formulation capable of simulating breaking waves impacting on tank walls, whereas the SPH results tended 
to overestimate the pressure maxima and impulse compared to experimental data. This may be caused by not 
considering the air phase in the numerical process. In addition, a considerable amount of work has demonstrated that 
the baffles significantly reduce the sloshing pressure acting on the walls on the basis of experiment and 
simulation results \cite{kim2001numerical, hwang2016numerical, al2021dynamic, trimulyono2022investigation}.
An experiment was carried out by Akyildiz and Unal \cite{akyildiz2005experimental} to study the non-linear 
behaviour and damping characteristics of liquid sloshing in the tank with/without baffles. 
Different fill depths and excitations were imposed on the sloshing model to investigate the pressure 
distributions and three-dimensional effects on sloshing loads. 

The displacement of the free surface stands as a crucial element when evaluating the sloshing intensity, and 
its tracking is consistently a fundamental challenge within tank sloshing simulations. Various computational 
methods in the Eulerian approach such as volume-of-fluid 
(VOF) \cite{chen2005time, nimisha2022effective, george2020anti} and level-set \cite{fang2007numerical} were 
introduced to address the discontinuity of liquid motion or breaking waves problems, which are often 
encountered in conventional grid-based methods. However, it is still a great challenge for them to handle 
large deformations of the fluid \cite{aly2015numerical}. In comparison, the boundary capture in both 
Lagrangian and arbitrary Lagrangian-Eulerian (ALE) approaches is more straightforward because the mesh 
precisely moves with fluid particles. 
As a meshless method, SPH offers remarkable benefits in addressing large deformations, 
free surfaces and moving boundaries due to its inherent self-adaptive and Lagrangian characteristics 
\cite{li2002meshfree, ye2019smoothed}. SPH has already been utilized in free-surface tracking in liquid 
sloshing problems by some 
researchers \cite{shao2012improved, cao2014sloshing, zhang2021improved, ren2023numerical}. 
Solving governing equations with a single-phase model in SPH can lead to overestimated 
pressure \cite{delorme2009set, pilloton2022sph} as mentioned earlier, whereas the accuracy of impact pressure simulations near 
the tank corners was confirmed to be enhanced in terms of mean and peak pressures by introducing the air phase 
to SPH simulation \cite{trimulyono2019experimental}. Yet now to the best of our knowledge, there is a scarcity 
of computational solutions addressing the deformable structures of tanks. Even though the force of fluid on 
the walls has been calculated in the published work, the detailed stress and strain distribution of the solid 
walls cannot be obtained from the numerical models employing rigid bodies. Therefore, it is in great demand 
for precise modelling of the interaction between the two-phase flow and the tank walls.

For the fluid-structure interaction problems, combining a single fluid solver with a single solid solver is 
commonly applied. FEM, as the most conventional methodology for solid mechanics, is often applied to treat 
structural deformations in FSI problems. In the previous experience in sloshing simulations, it has been 
coupled with grid-based methods such as FDM \cite{liao2013coupled, paik2014fluid} and FVM, or meshless methods 
such as moving particle semi-implicit (MPS) \cite{zhang2021partitioned} and SPH \cite{yang2012free}. 
Zhang et al. \cite{zhang2020investigations, zhang2021improved} proposed a coupling strategy involving the 
smoothed finite element method (SFEM) and an improved SPH to investigate sloshing mitigation using elastic 
baffles. In this strategy, SFEM excels in addressing the "overly-stiff" problem compared to the conventional 
FEM. Additionally, several computational efforts referring to liquid sloshing problems only rely on particle-based FSI solvers. 
Hwang \cite{hwang2016numerical} et al. presented a modified MPS-based FSI solver with the incorporation of a 
free-surface assessment scheme in the fluid model to enhance its performance. 
Hu et al. \cite{hu2019numerical} simulated the water in a rectangular tank interacting with an elastic baffle 
through the SPH-SPIM coupled method. However, the use of multiple solvers in a simulation can result in 
time-consuming, data loss or interpolation errors in the process of data transfer.

The objective of this work is to simulate the fluid flow regimes as well as the structural mechanics of walls 
during the sloshing procedure of LNG tanks, and to find out the damping mechanism of baffles on sloshing. 
Aiming at the above problems, we employ a multi-physics solver called SPHinXsys \cite{zhang2021sphinxsys}, 
which operates within a single framework. In this work, we first simulate both the fluid and solid domains 
using the SPH method. This approach avoids the excessive mesh distortion of fluid elements near the 
free-surface region in the grid-based methods, and also reduces the errors while transferring physical 
information between the fluid and solid domains. Second, the multi-phase model in terms of the water and air 
phases is conducted to be close to the actual utilization scenario of the LNG tanks. Third, the walls of the 
tanks are considered to be elastic to acquire the stress and strain distributions. All simulations are 
carried out in three dimensions to represent the actual physical fields better. The framework of this work 
is as follows: Section \ref{section: methodology} introduces the numerical methods in this work.
Section \ref{section: validation} provides the verification process to assess the reliability of the numerical 
simulation tool; Section \ref{section: elastic-wall} discusses the influence of ring baffles and vertical 
baffles on liquid sloshing degree and force exerted on tanks and Section \ref{section: rigid-elastic-baffle} 
compares the different effects of rigid and elastic baffles on sloshing suppression. 
Finally, some conclusions are drawn in Section \ref{section: conclusions}.

\section{Methodology} \label{section: methodology}
\subsection{Fluid dynamics method based on a multi-phase Riemann solver}

In this work, a multi-phase model consisting of two viscous and immiscible fluids with a high density ratio 
is adopted. The mass and momentum conservation equations can be expressed respectively as
\begin{equation}\label{eq: mass}
    \frac{\mathrm{d}\rho}{\mathrm{d}t} = -\rho \nabla \cdot \mathbf{v}
\end{equation}

\begin{equation}\label{eq: momentum}
    \frac{\mathrm{d}\mathbf{v}}{\mathrm{d}t} = \frac{1}{\rho} (-\nabla p + \eta \nabla^2 \mathbf{v}) 
    + \mathbf{f}^{s:p} + \mathbf{f}^{s:\nu} + \mathbf{f}^e
\end{equation}
where d/d$t$ is the material derivative, and $\rho$, $\mathbf{v}$, $p$, $\eta$ denote the density, the 
velocity, the pressure and the dynamic viscosity respectively. The force exerted on the fluid by the wall 
boundary is divided into pressure force $\mathbf{f}^{s:p}$ and viscous force $\mathbf{f}^{s:\nu}$. 
Besides, an external force $\mathbf{f}^e$ is added to the fluid.

The weakly compressible SPH (WCSPH) method is the most widely used method for incompressible multi-phase flows, 
while it may produce spurious pressure oscillation problems \cite{antuono2012numerical}. Aiming at the violent 
water-air flow cases in this work, the modified form of WCSPH in Ref.\cite{rezavand2020weakly} has been 
conducted. The modification refers to the low-dissipation Riemann solver and the application of 
transport-velocity formulation in the light phase. The fluid pressure is calculated by density from an 
artificial equation of state:
\begin{equation}\label{eq: eos}
    p = c^2 (\rho - \rho_0)
\end{equation}
where $c$ represents the numerical speed of sound and is defined as $c = 10V_{\rm max}$. 
The speed of sound of the heavy and light phases are the same. 
$V_{\rm max}$ is evaluated by $V_{\rm max} = 2\sqrt{gh_w}$, meaning the maximum anticipated flow speed. 
$g$ is the gravitational acceleration, $h_w$ is the depth of water and $\rho_0$ is the reference density 
of the fluid.

\begin{figure}
	\centering
	\begin{subfigure}[b]{0.35\textwidth}
		\centering
		\includegraphics[width=\textwidth]{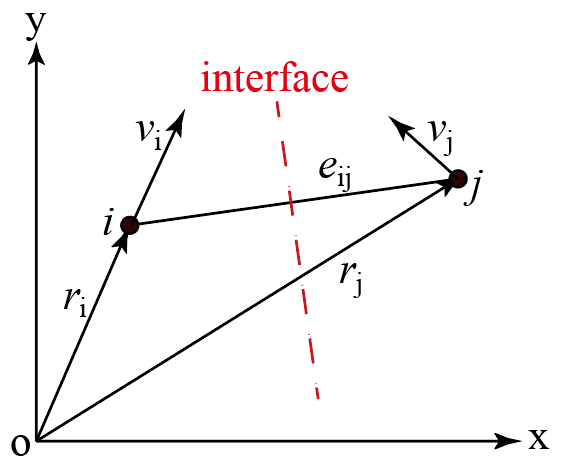}
		\caption{}
		\label{fig: riemann-a}
	\end{subfigure}
	\begin{subfigure}[b]{0.4\textwidth}
		\centering
		\includegraphics[width=\textwidth]{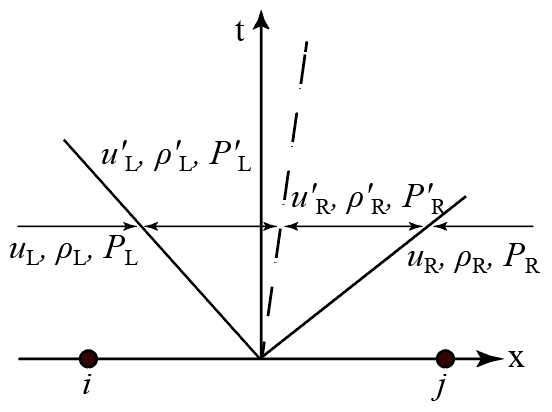}
		\caption{}
		\label{fig: riemann-b}
	\end{subfigure}
	\caption{WCSPH method based on a Riemann solver \cite{zhang2017weakly}: (a) Construction of Riemann 
 problem along the interacting line of particles i and j, (b) Simplified Riemann fan with two intermediate 
 states.}
	\label{fig: wcsph-riemann}
\end{figure}

In the WCSPH based on a multi-phase Riemann solver, the heavy phase is regarded as the moving boundary for 
the light phase, while itself is considered as a free-surface-like flow with variable free-surface pressure 
\cite{hu2004interface}. The SPH discretization of continuity and momentum equations can be written as:
\begin{equation}\label{eq: continuity-sph}
    \frac{\mathrm{d}\rho_i}{\mathrm{d}t} = 2\rho_i\sum_{j}\frac{m_j}{\rho_j}(\mathbf{v}_i - 
    \mathbf{v}^\ast)\cdot\nabla_iW_{ij}
\end{equation}

\begin{equation}\label{eq: momentum-sph}
    \frac{\mathrm{d}\mathbf{v}_i}{\mathrm{d}t} = -2\sum_{j}m_j \frac{p^\ast}{\rho_i\rho_j}\nabla_iW_{ij} + 
    2\sum_{j}m_j\frac{\eta_{ij}}{\rho_i\rho_j}
    \frac{\mathbf{v}_{ij}}{r_{ij}}\frac{\partial W_{ij}}{\partial r_{ij}} + \mathbf{f}_i^{s:p} + 
    \mathbf{f}_i^{s:\nu} + \mathbf{f}_i^e
\end{equation}
Here, $m_j$ is the particle mass, 
$\mathbf{v}_{ij} = \mathbf{v}_i - \mathbf{v}_j$ and $\eta_{ij} = \frac{2\eta_i\eta_j}{\eta_i + \eta_j}$. 
The gradient of kernel function $\nabla_iW_{ij} = \mathbf{e}_{ij} \frac{\partial W_{ij}}{\partial r_{ij}}$ 
with $\mathbf{e}_{ij} = -\mathbf{r}_{ij} / r_{ij}$ representing the unit vector from particle $i$ to $j$. 
$\mathbf{v}^\ast$ and $p^\ast$ denote the solutions of an inter-particle Riemann problem along 
$\mathbf{e}_{ij}$. The construction along the interface between particle $i$ and particle $j$ as well as the 
structure of the solution are shown in Fig.\ref{fig: wcsph-riemann}, with the initial left and right states 
reconstructed as
\begin{equation}\label{eq: wcsph-riemann}
\left\{\begin{aligned}
    & (\rho_L, U_L, p_L) = (\rho_i, \mathbf{v}_i \cdot \mathbf{e}_{ij}, p_i) \\
    & (\rho_R, U_R, p_R) = (\rho_j, \mathbf{v}_j \cdot \mathbf{e}_{ij}, p_j) \\
\end{aligned}\right.
\end{equation}
where subscripts $L$ and $R$ represent the left and right states of the one-dimensional Riemann problem. 
The middle wave is always a contact discontinuity separating two intermediate states, 
i.e. $(\rho_L^\ast, U_L^\ast, p_L^\ast)$ and $(\rho_R^\ast, U_R^\ast, p_R^\ast)$. 
An assumption $U^\ast = U_L^\ast = U_R^\ast$ and  $p^\ast = p_L^\ast = p_R^\ast$ is adopted at the intermediate 
state, and then the intermediate velocity and pressure can be approximated by
\begin{equation}\label{eq: velocity-pressure-riemann}
\left\{\begin{aligned}
    & U^\ast = \overline{U} + \frac{p_L - p_R}{c(\rho_L+\rho_R)} \\
    & p^\ast = \overline{p} + \frac{\rho_L\rho_R\beta(U_L-U_R)}{\rho_L+\rho_R} \\
\end{aligned}\right.
\end{equation}
Here, $\overline{U} = (\rho_LU_L + \rho_RU_R) / (\rho_L+\rho_R)$ and 
$\overline{p} = (\rho_Lp_R + \rho_Rp_L) / (\rho_L+\rho_R)$. 
The dissipation limiter $\beta = {\rm min}(3{\rm max}(U_L-U_R, 0), c)$ proposed in ref\cite{zhang2017weakly} 
is introduced for $p^\ast$ to reduce the numerical dissipation. 
Then $\mathbf{v}^\ast$ in Eq.(\ref{eq: momentum}) is derived by 
$\mathbf{v}^\ast = U^\ast\mathbf{e}_{ij} + (\overline{\mathbf{v}}_{ij} - \overline{U}\mathbf{e}_{ij})$, 
and $\overline{\mathbf{v}}_{ij} = (\rho_i\mathbf{v}_i + \rho_j\mathbf{v}_j) / (\rho_i+\rho_j)$.

Additionally, in fluid dynamics simulations, the tensile instability in the WCSPH method results in particle 
clumping and unnatural void regions when negative pressure occurs. Hence, the transport-velocity formulation 
presented by Adami et al. \cite{adami2013transport} and Zhang et al. \cite{zhang2017generalized} is applied to 
the light phase to avoid this issue. However, the heavy phase does not need it because the pressure 
distribution is always positive as a consequence of gravity. The particle advection velocity 
$\widetilde{\mathbf{v}}$ is obtained by employing a background pressure $p_b$ at every time step:
\begin{equation}\label{eq: advection-velocity}
    \widetilde{\mathbf{v}}_i(t+\delta t) = \mathbf{v}_i(t) + 
    \delta t(\frac{\mathrm{d}\mathbf{v}_i}{\mathrm{d}t} - 2p_b\sum_{j}m_j\frac{1}{\rho_i\rho_j}\nabla_iW_{ij})
\end{equation}
With the transport velocity, the momentum equation of Eq.\ref{eq: momentum-sph} can be rewritten as
\begin{equation}\label{eq: momentum-transportV}
    \frac{\mathrm{d}\mathbf{v}_i}{\mathrm{d}t} = -2\sum_{j}m_j \frac{p^\ast}{\rho_i\rho_j}\nabla_iW_{ij} 
    + 2\sum_{j}m_j\frac{\eta_{ij}}{\rho_i\rho_j}
    \frac{\mathbf{v}_{ij}}{r_{ij}}\frac{\partial W_{ij}}{\partial r_{ij}} 
    + 2\sum_{j}m_j\frac{\overline{\mathbf{A}}_{ij}}{\rho_i\rho_j}\nabla_iW_{ij} + \mathbf{g}_i
    + \mathbf{f}_i^{s:p} + \mathbf{f}_i^{s:\nu} + \mathbf{f}_i^e
\end{equation}
where $\overline{\mathbf{A}}_{ij} = (\mathbf{A}_i + \mathbf{A}_j) / 2$. 
$\mathbf{A}_i = \rho_i\mathbf{v}_i(\widetilde{\mathbf{v}}_i - \mathbf{v}_i)$ means an extra stress tensor due 
to the advection velocity $\widetilde{\mathbf{v}}$. 
Specifically, $\widetilde{\mathbf{v}}$ equals to $\mathbf{v}_j$ for the heavy phase.

\begin{figure}
  \centering
  {\includegraphics[width=3in]{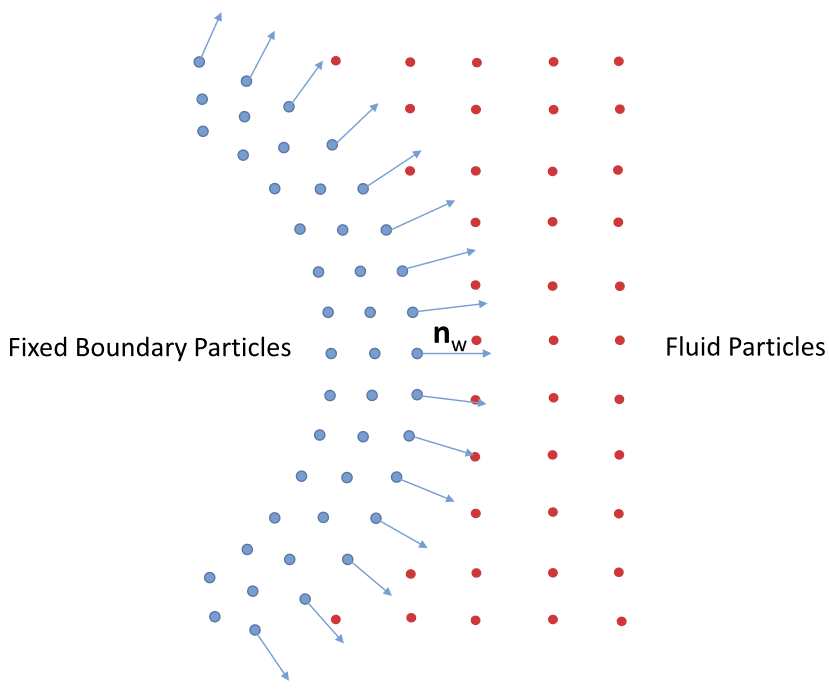}}
  \caption{\label{fig: one-sided-riemann}{Sketch of fluid particles interacting with fixed wall boundary 
  particles along the wall-normal direction through the one-side Riemann problem \cite{zhang2017weakly}.}}
\end{figure}

In SPHinXsys, dummy particles are adopted to impose the solid wall boundary \cite{adami2012generalized}. 
The interaction between the fluid particles and wall dummy particles is determined by the one-sided 
Riemann problem as shown in Fig.\ref{fig: one-sided-riemann} with the following intermediate pressure 
\cite{rezavand2022generalized}:
\begin{equation}\label{eq: one-sided-riemann-pressure}
    p^\ast = \frac{\rho_fp_w + \rho_wp_f}{\rho_f+\rho_w}
\end{equation}
where subscripts $f$ and $w$ represent the fluid and wall respectively. Also, $U^\ast$ is still acquired 
by Eq.(\ref{eq: velocity-pressure-riemann}). The pressure of wall dummy particles is calculated by
\begin{equation}\label{eq: one-sided-riemann-pressure-wall}
    p_w = \frac{\sum_{f}p_fW_{wf} + 
    (\mathbf{g} - \mathbf{a}_w)\cdot\sum_{f}\rho_f\mathbf{r}_{wf}W_{wf}}{\sum_{f}W_{wf}}
\end{equation}
where $a_w$ is the acceleration of the wall. Particularly, the Eq.(\ref{eq: one-sided-riemann-pressure-wall}) 
is divided by $\rho_f$ to remove the contribution of heavy phase particles in $p_w$ at a triple point, 
i.e. light phase, heavy phase and solid wall.

Besides, for the physical situations in the present study, penalty force is introduced to effectively prevent 
particle penetration in violent impacts and multi-phase flows with a high density ratio 
\cite{rezavand2022generalized}. When the particle of the light phase is close to the wall boundary, 
the penalty force can be calculated as
\begin{equation}\label{eq: penalty-force}
    \mathbf{F}_p = 
    -2\sum_{w}\frac{m_w}{\rho_i\rho_w}\Gamma(\mathbf{r}_i, \mathbf{r}_w)\mathbf{n}_w\frac{1}
    {|\mathbf{r}_{iw}|^2}\frac{\partial W_{iw}}{\partial (|\mathbf{r}_{iw}|)}
\end{equation}
Here subscripts $i$ and $w$ denote the light phase particles and wall particles, and $n_w$ is the predicted 
unit normal vector of the wall particles. $\Gamma(\mathbf{r}_i, \mathbf{r}_w)$ is the penalty parameter 
evaluated by
\begin{equation}\label{eq: penalty-parameter}
    \Gamma(\mathbf{r}_i, \mathbf{r}_w) = 
    \gamma_p|p_i\mathbf{e}_{iw}\cdot\mathbf{n}_w|\left\{\begin{aligned}
    & \frac{(1-\delta)^2}{\Delta x},& \delta < 1\\
    \ & 0, & \delta \ge 1
    \end{aligned}\right.
\end{equation}
where $\Delta x$ is the initial particle spacing and 
$\delta = 2|\mathbf{r}_{iw}|(\mathbf{e}_{iw}\cdot\mathbf{n}_w)/\Delta x$.

To further enhance the robustness and accuracy of the simulation approach, a reinitialization method referring 
to the density field is introduced in Ref.\cite{rezavand2022generalized}, which can be expressed as the 
following equations for the free-surface-like flow of the water as well as the air with the moving boundary 
respectively.
\begin{equation}\label{eq: water-density-reinitialization}
    \rho_i = \rho^0\frac{\sum W_{ij}}{\sum W_{ij}^0} + 
    {\rm max}(0, \rho^\ast-\rho^0\frac{\sum W_{ij}}{\sum W_{ij}^0})\frac{\rho^0}{\rho^\ast}
\end{equation}

\begin{equation}\label{eq: air-density-reinitialization}
     \rho_i = \rho^0\frac{\sum W_{ij}}{\sum W_{ij}^0}
\end{equation}
where $\rho^\ast$ stands for the density before reinitialization and the subscript 0 means the initial state.

\subsection{Solid Dynamics}

The deformation tensor $\mathbb{F}$ can be derived from the displacement 
$\mathbf{u} = \mathbf{r} - \mathbf{r}^0$ by
\begin{equation}
     \mathbb{F} = \nabla^0\mathbf{u} + \mathbf{I}
\end{equation}

The mass and momentum conservation equations of solid mechanics in a Lagrangian framework are established as
\begin{equation}
     \rho=\rho_0\frac{1}{J}
\end{equation}

\begin{equation}
     \frac{\mathrm{d}\mathbf{v}}{\mathrm{d}t} = \frac{1}{\rho^0}\nabla^0\cdot\mathbb{P}^T + 
     \mathbf{f}^{f:p} + \mathbf{f}^{f:\nu}
\end{equation}
where $J = det(\mathbb{F})$ is the Jacobian determinant of deformation tensor $\mathbb{F}$, 
$\mathbb{P} = \mathbb{F}\mathbb{S}$ is the first Piola-Kirchhoff stress tensor with $\mathbb{S}$ denoting the 
second Piola-Kirchhoff stress tensor, which can be simplified by Eq.(\ref{eq19}) when the material is linearly 
elastic and isotropic. $\mathbf{f}^{f:p}$ and $\mathbf{f}^{f:\nu}$ represent the pressure and viscous forces 
from the fluid on the solid.
\begin{equation}\label{eq19}
\begin{split}
     \mathbb{S} & = 
     K {\rm tr}(\mathbb{E}) \mathbf{I} + 2G(\mathbb{E} - \frac{1}{3} {\rm tr}(\mathbb{E})\mathbf{I})\\
     & = \lambda {\rm tr}(\mathbb{E}) \mathbf{I} + 2 \mu \mathbb{E}
\end{split}
\end{equation}
where $K = \lambda + (2\mu/3)$ and $G = \mu$ represent the bulk modulus and shear modulus respectively. 
$\lambda$ and $\mu$ are the Lam$\Acute{\rm e}$ parameters. Note that the bulk and shear modulus can be 
expressed by the Young's modulus $E$ and Poisson's ratio $\nu$ as Eq.(\ref{eq: Young-modulus}). 
The Green-Lagrange strain tensor $\mathbb{E}$ is described based on the deformation tensor 
$\mathbb{F}$ by Eq.(\ref{eq: strain-tensor}).
\begin{equation}\label{eq: Young-modulus}
     E = 2G(1+2\nu) = 3K(1-2\nu)
\end{equation}

\begin{equation}\label{eq: strain-tensor}
     \mathbb{E} = \frac{1}{2} (\mathbb{F}^T\mathbb{F} - \mathbf{I})
\end{equation}

SPHinXsys adopts the total Lagrangian formulation for solid mechanics. The initial reference configuration is 
applied to search the neighboring particles and the set of neighboring particles is not updated. 
The mass and momentum conservation equation for solid particles are discretized as
\begin{equation}\label{eq: solid-density}
    \rho_i = \rho^0 \frac{1}{\rm det(\mathbb{F})}
\end{equation}

\begin{equation}\label{eq: solid-momentum}
    \frac{\mathrm{d}\mathbf{v}_a}{\mathrm{d}t} = 
    \sum_{b} \frac{m_b}{\rho_a\rho_b}(\mathbb{P}_a\mathbb{B}_a^0+\mathbb{P}_b\mathbb{B}_b^0)
    \nabla_a^0W_{ab} + \mathbf{f}_a^{s:p} + \mathbf{f}_a^{s:\nu}
\end{equation}
where the subscript $a$ means the solid particle. The correction matrix 
$\mathbb{B}^0_a = 
\big{(}\sum_{b}\frac{m_b}{\rho_b}(\mathbf{r}^0_b-\mathbf{r}^0_a)\otimes \nabla_a^0W_{ab} \big{)^{-1}}$ 
is introduced to ensure the homogeneity and isotropy of space in the process of 
spatial discretization \cite{vignjevic2006sph}. The deformation tensor $\mathbb{F}$ is calculated as
\begin{equation}\label{eq: deformation-tensor}
    \mathbb{F} = 
    \bigg{(}\sum_{b}\frac{m_b}{\rho_b}(\mathbf{u}_b-\mathbf{u}_a)\otimes \nabla_a^0W_{ab} \bigg{)}\mathbb{B}_a^0 + \mathbf{I}
\end{equation}

\subsection{Fluid-Structure Interaction}

As mentioned in the introduction section, the fluid-structure interaction among water, air and wall will be 
conducted in this paper. SPHinXsys benefits from its multi-resolution framework, namely that the fluid and 
solid domains in FSI problems can be discretized by different spatial-temporal resolutions 
\cite{zhang2021multi}. The smoothing length for fluid and solid discretization are expressed 
as $h_f$ and $h_s$, and $h_f \ge h_s$. In this work, $h_f = 1.3\Delta x$ and $h_s = 1.15\Delta x$. 
Detailedly, the pressure force and viscous force exerted on fluid by solid in 
Eq.(\ref{eq: momentum-transportV}) are described as
\begin{equation}\label{eq: fsi-pressure-on-fluid}
    \mathbf{f}^{s:p}_i(h_f) = -2\sum_{a}m_a\frac{p^\ast}{\rho_i\rho_a}\nabla_iW(\mathbf{r}_{ia}, h_f)
\end{equation}

\begin{equation}\label{eq: fsi-viscous-on-fluid}
    \mathbf{f}^{s:\nu}_i(h_f) = 
    2\sum_{a}m_a\frac{\eta_{ia}}{\rho_i\rho_a}\frac{\mathbf{v}_i - \mathbf{v}_a^d}{|\mathbf{r}_{ia}| 
    + 0.01h}\frac{\partial W(\mathbf{r}_{ia}, h_f)}{\partial r_{ia}}
\end{equation}
Here, $p^\ast =(p_i\rho_a^d + p_a^d\rho_i)/ (\rho_i+\rho^d_a)$ is the Riemann solution of the one-sided 
Riemann problem whose left and right states are given as Eq.(\ref{eq: fsi-riemann-solution}). $\mathbf{v}_d$ 
is the imaginary velocity of the solid particle. 
\begin{equation}\label{eq: fsi-riemann-solution}
\left\{\begin{aligned}
    & (\rho_L, U_L, p_L) = (\rho_i, -\mathbf{v}_i \cdot \mathbf{n}_{a}, p_i) \\
    & (\rho_R, U_R, p_R) = (\rho_a, -\mathbf{v}_a^d \cdot \mathbf{n}_{a}, p_a^d) \\
\end{aligned}\right.
\end{equation}
where $\mathbf{n}_a$ means the local normal vector from solid particle to fluid particle. 
The forces acting on the solid by fluid are described as
\begin{equation}\label{eq: forces-on-solid}
\left\{\begin{aligned}
    & \mathbf{f}^{f:p} = - \mathbf{f}^{s:p} \\
    & \mathbf{f}^{f:\nu} = - \mathbf{f}^{s:\nu}. \\
\end{aligned}\right.
\end{equation}

\subsection{Sloshing Making}

In the present work, pitching excitation is considered to simulate the liquid sloshing in the tank. 
The tank is assumed to be fixed and the reverse external force is applied to the multi-phase flow. 
According to Xue and Lin \cite{xue2011numerical}, the external acceleration $\mathbf{f}^e$ of the
momentum equation in Eq.(\ref{eq: momentum}) for both water and air phases are expressed as
\begin{equation}\label{eq: external-acc}
    \mathbf{f} = 
    \mathbf{g} - \frac{\mathrm{d}\mathbf{U}}{\mathrm{d}t} - 
    \frac{\mathrm{d}\mathbf{\Omega}}{\mathrm{d}t} \times (\mathbf{r} - 
    \mathbf{R}) - 
    2\mathbf{\Omega} \times \frac{\mathrm{d}\mathbf{\mathbf({r} - 
    \mathbf{R})}}{\mathrm{d}t} - \mathbf{\Omega}\times[\mathbf{\Omega}\times(\mathbf{r} - \mathbf{R})]
\end{equation}
where $\mathbf{U}$ and $\mathbf{\Omega}$ are the translational and rotational velocities of non-inertial 
coordinate respectively. $\mathbf{R}$ and $\mathbf{r}$ are the original point of rotational motion and 
the local position vector.

Specifically, to prevent the tank from running away under the force of liquid when the walls of the tank are 
replaced with elastic material, the constraints composed of translation and rotation are proposed for fixing 
the tank. The total momentum of the tank $\mathbf{P}$ is calculated through
\begin{equation}
    \mathbf{P} = \sum_{a} m_a \cdot \mathbf{v}_a
\end{equation}

Subsequently, the translational velocity of the tank $\mathbf{v}_{\rm tank}$ is computed through the 
equation $\mathbf{v}_{\rm tank} = \mathbf{P}/m_{\rm total}$ with the total mass of tank $m_{\rm total}$. 
Further, the velocity of each tank particle should be subtracted from the translational velocity. 
Once the tank's translational movement is confined, the rotational motion will conduct around its 
center of mass The angular momentum of the tank $\mathbf{L}$ is written as
\begin{equation}
    \mathbf{L} = \mathbf{r} \times (m\mathbf{v}) = \mathbb{I} \mathbf{\omega}
\end{equation}

Therefore, the angular velocity of the tank $\mathbf{\omega}$ equals to $\mathbb{I}^{-1} \times \mathbf{L}$, 
where $\mathbb{I}$ signifies the inertia tensor. Then the velocity of each tank particle should be adjusted 
by subtracting the linear velocity of the tank, 
i.e. $\mathbf{v}^{\rm linear}_{\rm tank} = \omega \times \mathbf{r}$.

\subsection{Artificial-viscosity-based damping method}

To initiate sloshing while the water is relatively still, only downward gravitational acceleration is 
applied to the fluid during the initial second. Moreover, an extra artificial-viscosity-based damping term, 
as described in Ref.\cite{zhu2022dynamic}, is introduced into the momentum equation at this stage to 
effectively dissipate velocity gradients while preserving the system's momentum conservation. 
The local damping term in implicit format can be written as
\begin{equation}\label{eq: viscous-damping}
    \mathbf{f}^{\nu}_i = 
    2\eta\sum_{j}\frac{m_j}{\rho_i\rho_j}\frac{\mathbf{v}_{ij}^{n+1}}{\mathbf{r}_{ij}}
    \frac{\partial W_{ij}}{\partial r_{ij}}
\end{equation}
where $\mathbf{v}_{ij}^{n+1} = \mathbf{v}_{ij}^{n} + \mathrm{d}\mathbf{v}_i - \mathrm{d}\mathbf{v}_j$ with 
$\mathrm{d}\mathbf{v}_i$ and $\mathrm{d}\mathbf{v}_j$ denoting the incremental velocity change of 
particle $i$ and its neighboring particles $j$ (including both internal and contact configurations of the 
particle $i$) induced by viscous acceleration.

Further, here we employ the pairwise splitting scheme proposed by Zhu\cite{zhu2022dynamic} for an implicit 
velocity update, from which $\mathrm{d}\mathbf{v}_i$ and $\mathrm{d}\mathbf{v}_j$ can be derived as follows:
\begin{equation}\label{eq: damping-velocity-increment}
\left\{\begin{aligned}
    & \mathrm{d}\mathbf{v}_i = m_j\frac{B_j\mathbf{v}_{ij}}{m_im_j-(m_i+m_j)B_j} \\
    & \mathrm{d}\mathbf{v}_j = -m_i\frac{B_j\mathbf{v}_{ij}}{m_im_j-(m_i+m_j)B_j} \\
\end{aligned}\right.
\end{equation}
where $B_j$ is defined as
\begin{equation}\label{eq: B-equation}
    B_j = 2\eta\frac{m_j}{\rho_i\rho_j}\frac{1}{\mathbf{r}_{ij}}\frac{\partial W_{ij}}{\partial r_{ij}}\mathrm{d}t.
\end{equation}

\subsection{Time Steps}

The time step size of fluid and solid simulation are limited by the CFL condition for numerical stability.
The dual-criteria time stepping for fluid dynamics is employed in SPHinXsys, which is characterized by 
particle advection and acoustic speeds \cite{zhang2020dual}. More precisely, the time-step size determined by 
the advection criterion, symbolized as $\Delta t_{ad}$, and that determined by the acoustic criterion, 
symbolized as $\Delta t_{ac}$, are defined through Eq.(\ref{eq: advection-time}) and 
Eq.(\ref{eq: acoustic-time}). The advection criterion controls the update of the neighbor particle list and the 
corresponding kernel weights and gradients, while the acoustic criterion determines the time integration of 
the particle density, position and velocity.
\begin{equation}\label{eq: advection-time}
    \Delta t_{ad} = CFL_{ad} {\rm min}(\frac{h}{|\mathbf{v}|_{\rm max}}, \frac{h^2}{\nu})
\end{equation}

\begin{equation}\label{eq: acoustic-time}
    \Delta t_{ac} = CFL_{ac} \frac{h}{c+|\mathbf{v}|_{\rm max}}
\end{equation}
where $CFL_{ad} = 0.25$ and $CFL_{ac} = 0.6|\mathbf{v}|_{\rm max}$ and $\nu$ mean the maximum particle 
advection speed and kinematic viscosity respectively.

For the time integration of solid equations, the time-step criterion $\Delta t^s$ is defined as
\begin{equation}\label{eq: solid-time}
    \Delta t^s = 0.6 {\rm min}(\frac{h^s}{c^s+|\mathbf{v}|_{\rm max}}, 
    \sqrt{\frac{h^s}{|\frac{\mathrm{d}\mathbf{v}}{\mathrm{d}t}|_{\rm max}}})
\end{equation}

Further, the structure time stepping is coupled with the dual-criteria time stepping for the FSI problem. 
In SPHinXsys, the position-based Verlet scheme is employed, that is in one advection time 
step $\Delta t_{ad}$, the acoustic time step $\Delta t_{ac}$ is repeated for the pressure relaxation process 
until the accumulated time interval exceeds $\Delta t_{ad}$. 
At the beginning of every advection time step, the density field is reinitialized by 
Eq.(\ref{eq: water-density-reinitialization}) and Eq.(\ref{eq: air-density-reinitialization}) for heavy and 
light phases respectively. The first half-step velocity in the $n$ acoustic time step is updated as
\begin{equation}\label{eq: midpoint-fluid-velocity}
    \mathbf{v}^{n+\frac{1}{2}}_i = 
    \mathbf{v}^{n}_i + \frac{\Delta t_{ac}}{2}(\frac{\mathrm{d}\mathbf{v}_i}{\mathrm{d}t})^n
\end{equation}
Then the updated velocity at the midpoint is applied to obtain the particle position and density in the 
meantime for the next acoustic time step
\begin{equation}\label{eq: update-fluid-position}
    \mathbf{r}^{n+1}_i = \mathbf{r}^{n}_i + \Delta t_{ac}\mathbf{v}_i^{n+\frac{1}{2}}
\end{equation}

\begin{equation}\label{eq: update-fluid-density}
    \rho^{n+1}_i = \rho^{n}_i + 
    \frac{\Delta t_{ac}}{2}(\frac{\mathrm{d}\rho_i}{\mathrm{d}t})^{n+\frac{1}{2}}
\end{equation}

At last, the velocity of the particle $i$ at the end of this acoustic time step is obtained by
\begin{equation}\label{eq: update-fluid-velocity}
    \mathbf{v}^{n+1}_i = 
    \mathbf{v}^{n}_i + \frac{\Delta t_{ac}}{2}(\frac{\mathrm{d}\mathbf{v}_i}{\mathrm{d}t})^{n+1}
\end{equation}

For the time integration of solid equations, generally $\Delta t^s < \Delta t_{ac}$. 
Index $x = 0, 1, ... k-1$ is utilized within one acoustic time step of fluid integration 
with $k = \big[\frac{\Delta t_{ac}}{\Delta t^s} \big ] + 1$. The deformation tensor, density and 
particle position are updated to the midpoint as
\begin{equation}\label{eq: midpoint-solid-deformation-tensor}
    \mathbb{F}^{x+\frac{1}{2}}_a = 
    \mathbb{F}^{x}_a + \frac{\Delta t^{s}}{2}\frac{\mathrm{d}\mathbb{F}_{a}}{\mathrm{d}t}
\end{equation}

\begin{equation}\label{eq: midpoint-solid-density}
    \rho_a^{x+\frac{1}{2}} = \rho^0_a \frac{1}{J}
\end{equation}

\begin{equation}\label{eq: midpoint-solid-position}
    \mathbf{r}_a^{x + \frac{1}{2}} = \mathbf{r}_a^x + \frac{\Delta t^{s}}{2} \mathbf{v}_a^x
\end{equation}

After that, the velocity of solid particle $a$ is updated to the next time step
\begin{equation}\label{eq: update-solid-velocity}
    \mathbf{v}^{x+1}_a = \mathbf{v}^{x}_a + \Delta t^s\frac{\mathrm{d}\mathbf{v}_{a}}{\mathrm{d}t}
\end{equation}

Finally, the deformation tensor and position of solid particles are updated to the new time step by
\begin{equation}\label{eq: update-solid-deformation-tensor}
    \mathbb{F}^{x+1}_a = \mathbb{F}^{x + \frac{1}{2}}_a 
    + \frac{\Delta t^{s}}{2}\frac{\mathrm{d}\mathbb{F}_{a}}{\mathrm{d}t}
\end{equation}

\begin{equation}\label{eq: update-solid-density}
    \rho_a^{x+1} = \rho^0_a \frac{1}{J}
\end{equation}

\begin{equation}\label{eq: update-solid-position}
    \mathbf{r}_a^{x+1} = \mathbf{r}_a^{x +\frac{1}{2}} + \frac{\Delta t^{s}}{2} \mathbf{v}_a^{x+1}
\end{equation}

\subsection{Algorithm}

The simulation procedure of the multi-phase liquid sloshing in an elastic tank is shown in Algorithm \ref{alg: 1}.

\newgeometry{left=2cm, right=2cm, top=3cm, bottom=3cm}

\begin{algorithm}[htb!]
    Setup geometrical and material parameters\;
    Generate particle distribution\;
    \While{simulation termination condition is not satisfied}
        {
        Calculate $\Delta t_{ad}$ for water and air\;
        $\mathrm{D}t = {\rm min}(\Delta t_{ad}^{w}, \Delta t_{ad}^{a})$\;
        Reinitialize density with Eq.(\ref{eq: water-density-reinitialization}) for water and
        Eq.(\ref{eq: air-density-reinitialization}) for air\;
        Calculate transport velocity for air with Eq.(\ref{eq: momentum-transportV})\;
        Calculate viscous force on fluid particles with viscosity term in Eq.(\ref{eq: fsi-viscous-on-fluid})\;
        Calculate viscous force on the tank wall with Eq.(\ref{eq: fsi-viscous-on-fluid} 
        and Eq.(\ref{eq: forces-on-solid})\;
        Update wall normal vector\;
        \While{relaxation time $<$ $\mathrm{D}t$}
            {
            Calculate $\Delta t_{ac}$ for water and air\;
            $dt = {\rm min}(\Delta t_{ac}^{w}, \Delta t_{ac}^{a}, \mathrm{D}t)$\;
            \While{physical time $<$ 1 second}
                {
                Execute fluid damping process with Eq.(\ref{eq: viscous-damping})\;
                }
            Update the velocity of fluid particles to the midpoint with Eq.(\ref{eq: midpoint-fluid-velocity})\;
            Update the pressure force on fluid particles with Eq.(\ref{eq: fsi-pressure-on-fluid}\;
            Update the pressure force on the tank wall with Eq.(\ref{eq: forces-on-solid})\;
            Update the position and density of fluid particles with Eq.(\ref{eq: update-fluid-position}) and
            Eq.(\ref{eq: update-fluid-density}) respectively\;
            Update the velocity of fluid particles to next time step with Eq.(\ref{eq: update-fluid-velocity})\;
            \While{$\mathrm{d}t^s_{\rm sum} < \mathrm{d}t$}
                {
                Calculate $\Delta t^s$ for solid\;
                $dt^s = {\rm min}(\Delta t^s, \mathrm{d}t - \mathrm{d}t^s_{\rm sum})$\;
                Update $\mathbb{F}$, $\rho$ and $\mathbf{r}$ of solid particles to the midpoint with 
                Eq.(\ref{eq: midpoint-solid-deformation-tensor}), Eq.(\ref{eq: midpoint-solid-density})
                and Eq.(\ref{eq: midpoint-solid-position})\;
                Update the velocity of solid particles with Eq.(\ref{eq: update-solid-velocity})\;
                Constrain the elastic tank\;
                Update $\mathbb{F}$, $\rho$ and $\mathbf{r}$ of solid particles to the next time step with 
                Eq.(\ref{eq: update-solid-deformation-tensor}), Eq.(\ref{eq: update-solid-density}) 
                and Eq.(\ref{eq: update-solid-position})\; 
                }
            Update neighbor particle list, kernel values and gradient\;
            State Update particle configuration\;
            }
        }
    Terminate the simulation.
    \caption{Multi-phase liquid sloshing in an elastic tank}
    \label{alg: 1}
\end{algorithm}
\restoregeometry 
\section{Validation Tests} \label{section: validation}
\subsection{Comparison with LNG tank experiment}

\begin{figure}
  \centering
  {\includegraphics[width=3in]{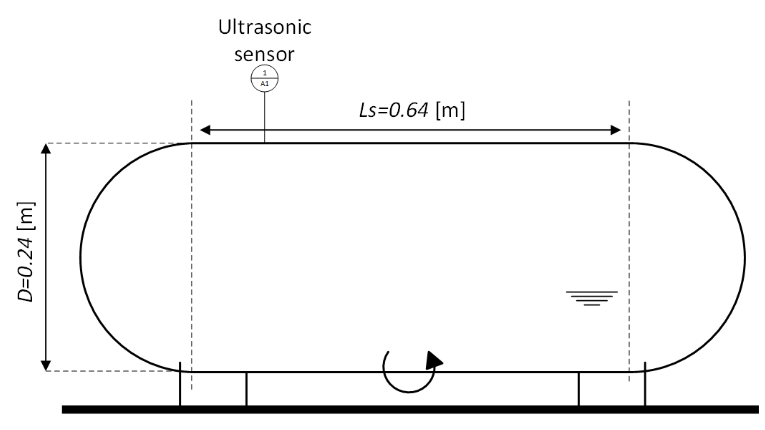}}
  \caption{\label{fig: experiment-setup}{Experimental test setup \cite{grotle2017experimental}}}
\end{figure}

The previous research that simplifies the tank as a rigid body is often insufficiently accurate. 
The complex interaction between the tank's motions and deformations substantially impacts the predicted 
accuracy of the strain and stress distribution of the walls and the flow characteristics of the fluid. 
In this case, the tank walls are modelled as elastic bodies to provide a more comprehensive and realistic 
simulation condition. The simulation results are compared with the experiment carried out 
by Grotle and Æsøy \cite{grotle2017experimental} to validate the prediction accuracy of SPHinXsys. 
The experimental setup is shown in Fig.\ref{fig: experiment-setup}, featuring a cylindrical tank with 
semi-elliptical heads at a 2:1 ratio, and the thickness of the tank is modelled as 0.018 meters. 
In this test, the tank undergoes pitching motion excitation, with the roll axis positioned at the 
intersection of the tank bottom and the center plane. The tank's motion angle is given as 
$\theta(t) \approx Asin(2 \pi ft)$, where $A$ is the constant amplitude equal to 3 degrees, 
$f$ [Hz] is the frequency of the motion with the value of 0.5496 and $t$ [s] represents the physical time. 
A sensor is located at 0.122m away from the left-hand beginning of the straight cylindrical section to detect 
the free-surface height. For the validation case, we adopt a ratio of 0.255 between the water depth and the 
inside diameter of the tank, denoted as $h_w/D$. Additionally, the densities of the water and air 
are 1000 kg/m$^3$ and 1.226 kg/m$^3$ respectively.

\begin{figure}
  \centering
  {\includegraphics[width=4in]{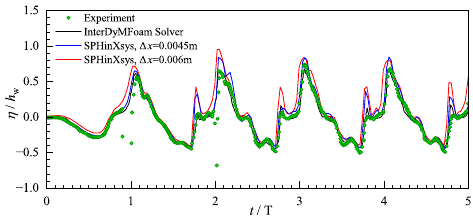}}
  \caption{\label{fig: validation}{Comparison of non-dimensional free surface elevation of water 
  between experiment \cite{grotle2017experimental} and simulations.}}
\end{figure}

Fig.\ref{fig: validation} indicates the comparison of the free surface elevation predicted by SPHinXsys, 
three-dimensional RANS model based on interDyMFoam solver with OpenFOAM and experimental data provided in 
Ref.\cite{grotle2017experimental}. Initial particle spacing values $\Delta x = 0.0045$ and 0.006 m are tested 
for grid independence, noting that the lower resolution does not provide three layers of wall particles required 
to achieve sufficient kernel support for fluid neighbors. The x-axis $t/\rm{T}$ stands for the ratio of physical 
time to the period of circular motion, while the y-axis $\eta/h_w$ means the ratio of free surface elevation to 
the initial water depth. It is noted that the numerical sloshing wave elevation predicted by SPHinXsys closely 
follows the experimental trend, and its accuracy improves with increasing resolution. The pattern and period 
of wave heights obtained by SPH simulation are comparable with experimental observations. However, SPH tends 
to overestimate the amplitude, although it remains within an acceptable range. This higher amplitude estimation 
could be attributed to the influence of some splashing water particles on wave height estimation. 
To balance computational efficiency and numerical reliability, $\Delta x = 0.006$ m is chosen for 
the subsequent calculation.

\begin{figure}
  \centering
  {\includegraphics[width=3in]{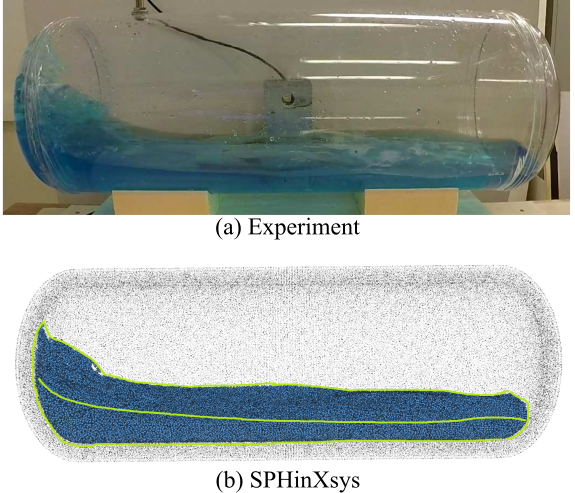}}
  \caption{\label{fig: wave-exp-vs-sph}{Wave profiles comparison between (a) experiment and (b) SPHinXsys.}}
\end{figure}

\begin{figure}
  \centering
  {\includegraphics[width=4in]{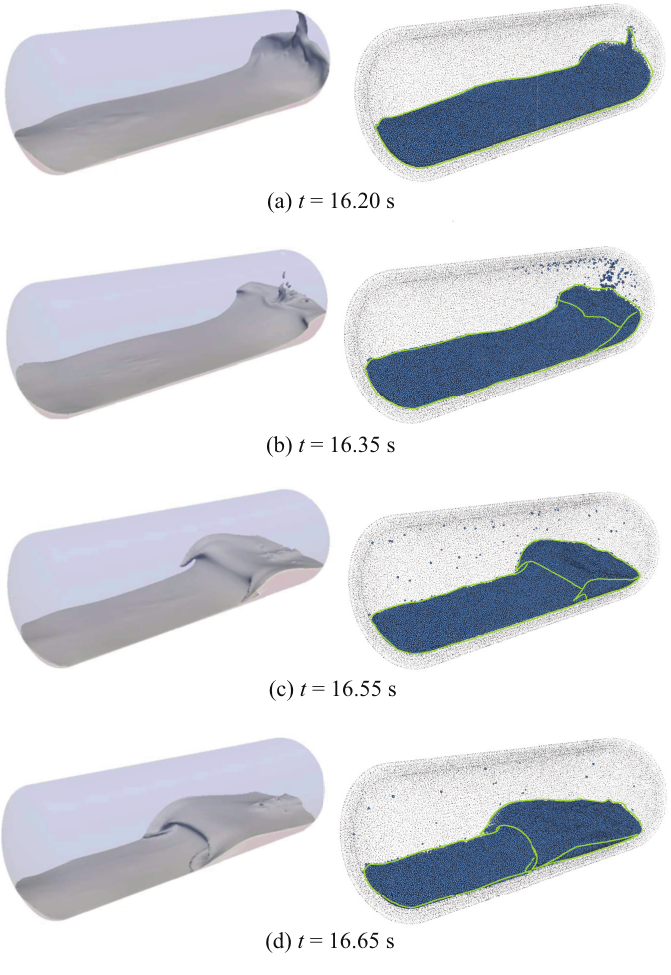}}
  \caption{\label{fig: wave-interDyMFoam-vs-sph}{Wave profiles at different time: 
  interDyMFoam-solver-based simulation (left-hand panels) \cite{grotle2017experimental} and SPHinXsys (right-hand panels).}}
\end{figure}

Fig.\ref{fig: wave-exp-vs-sph} presents the numerical and experimental free surface wave profiles when the 
wave travels back from the tank head. Fig.\ref{fig: wave-interDyMFoam-vs-sph} depicts the comparison of 
predicted wave profiles between interDyMFoam solver and SPHinXsys at several moments. 
Here the air domain is not displayed in SPH results for better visualization. 
At 16.20s, the wave reaches the frontal section of the tank, leading to the formation of a vertical fountain. 
The wave profile illustrated on the right-hand panels by SPH closely resembles that on the left-hand panels 
at the same time. 
Subsequently, as the wave retreats from the tank head, the vertical fountain dissipates. At this juncture, 
the wave profile obtained by SPH aligns remarkably well with the corresponding snapshot from the simulation 
conducted by the interDyMFoam solver. Following this, the wave maintains its propagation, prompting a rapid 
reduction in water depth near the wall. This observable phenomenon is captured in snapshots from both the 
SPHinXsys and interDyMFoam-solver-based simulations. Specifically, at 16.55s and 16.65s, 
the wave profiles simulated by SPHinXsys exhibit strong concordance with the results obtained through the 
interDyMFoam solver. The above analyses demonstrate that the SPHinXsys algorithm is effective in predicting 
liquid sloshing problems under pitching excitation.

\subsection{Verification work before} 

Another robust validation is presented in the published literature by Ren et al. using 
SPHinXsys \cite{ren2023numerical}. According to their work, the comparison between numerical results and 
experimental data or analysis solutions of liquid sloshing is performed, and satisfactory agreement can be 
obtained. This proves the reliability of SPHinXsys in evaluating parameters such as free surface elevation, 
pressure forces on baffles, and the displacement of elastic baffles. Therefore, the SPH method can be 
further used in the following work on free surface capture, force and deformation analysis referring to 
solid mechanics as well as the fluid-structure interaction process.

\section{Analysis of fluid sloshing and force on the tank with/without baffles} \label{section: elastic-wall}

This chapter involves the investigation of the force on the tank during the liquid sloshing process and 
the baffle effect on the suppression of sloshing behaviors. The density, Poisson ratio and Young's modulus 
of the tank are 7890 kg/m$^3$, 0.27 and 135 GPa respectively. Note that the analysis begins with the 
sloshing behavior, i.e. the preloading process with the damping method is ignored.

\begin{figure}
	\centering
	\begin{subfigure}[b]{0.3\textwidth}
		\centering
		\includegraphics[width=\textwidth]{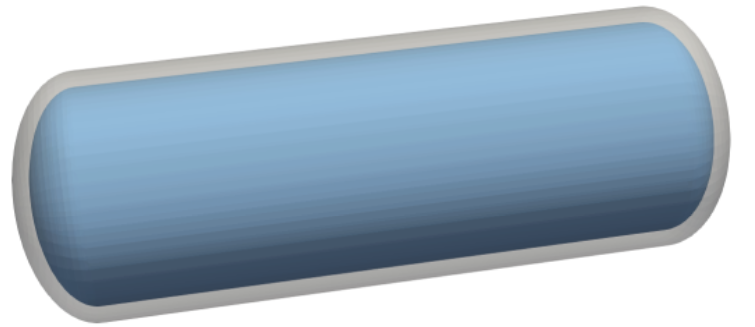}
		\caption{}
		\label{fig: basic-geo}
	\end{subfigure}
	\begin{subfigure}[b]{0.3\textwidth}
		\centering
		\includegraphics[width=\textwidth]{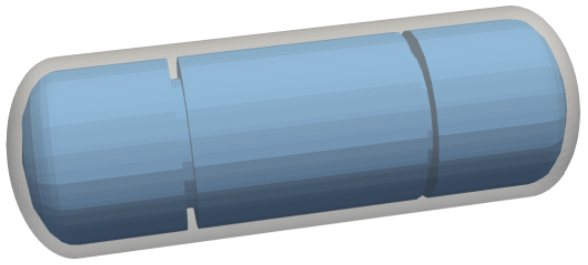}
		\caption{}
		\label{fig: ring-geo}
	\end{subfigure}
	\begin{subfigure}[b]{0.3\textwidth}
		\centering
		\includegraphics[width=\textwidth]{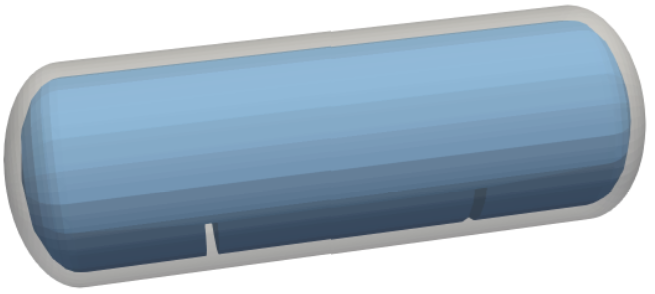}
		\caption{}
		\label{fig: vertical-geo}
	\end{subfigure}
	\caption{Structures of elastic tanks: (a) without baffle, (b) with 2 ring baffles and 
 (c) with 2 vertical baffles.}
	\label{fig: tank-structure}
\end{figure}

\begin{figure}
  \centering
  {\includegraphics[width=4in]{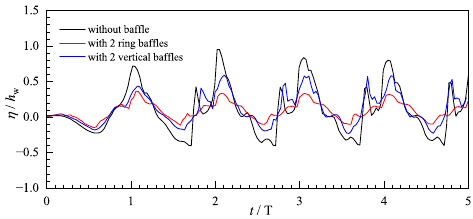}}
  \caption{\label{fig: ring-vertical-baffles}{Comparison of non-dimensional free surface elevation 
  of water among normal tank and tank with baffles}}
\end{figure}

\begin{figure}
  \centering
  {\includegraphics[width=5in]{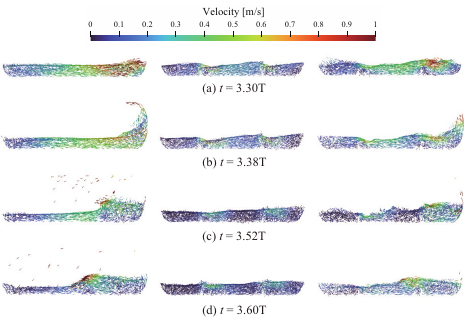}}
  \caption{\label{fig: basic-ring-vertical-velocity}{Velocity distributions of water in the third period. 
  left: elastic tank without baffle, middle: elastic tank with 2 ring baffles, right: elastic tank 
  with 2 vertical baffles.}}
\end{figure}

\begin{figure}
  \centering
  {\includegraphics[width=5in]{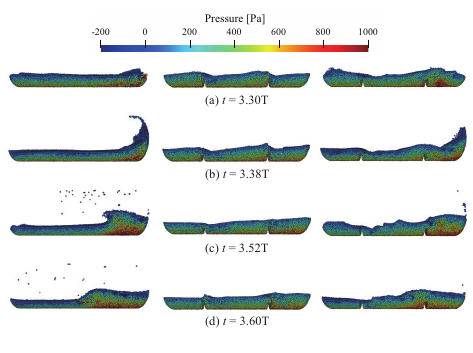}}
  \caption{\label{fig: basic-ring-vertical-pressure}{Pressure distributions of water in the third period. 
  left: elastic tank without baffle, middle: elastic tank with 2 ring baffles, right: elastic tank 
  with 2 vertical baffles.}}
\end{figure}

\begin{figure}
  \centering
  {\includegraphics[width=5in]{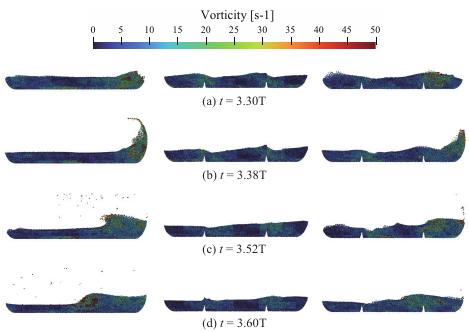}}
  \caption{\label{fig: basic-ring-vertical-vorticity}{Vorticity distribution of three typical flow patterns. 
  left: elastic tank without baffle, middle: elastic tank with 2 ring baffles, right: elastic tank 
  with 2 vertical baffles.}}
\end{figure}

Fig.\ref{fig: tank-structure} depicts the structures of the standard LNG tank as well as the tank equipped 
with two kinds of baffles: ring baffles and vertical baffles. Fig.\ref{fig: ring-vertical-baffles} presents 
a comparison of the non-dimensional free surface elevation of water among three tank configurations. 
The corresponding fluid regimes in the third period, when the sloshing becomes stable, are displayed in 
Fig.\ref{fig: basic-ring-vertical-velocity}, Fig.\ref{fig: basic-ring-vertical-pressure} and 
Fig.\ref{fig: basic-ring-vertical-vorticity}. Here, four representative moments are selected, 
including fluid preparing to swing to the right side, fluid hitting the right head of the tank and 
backflow after impact. Notably, the introduction of baffles facilitates mitigating sloshing effectively 
by disrupting wave propagation. In other words, the tanks incorporated two ring or vertical baffles alter 
the flow patterns of water and consequently affect the free surface height. Relatively speaking, 
ring baffles have a more pronounced effect on mitigating large oscillations in this numerical case.

The velocity of the primary wave occupies the maximum value, and the integration of baffles yields a 
noticeable refinement in the velocity distribution, resulting in a smoother pressure field and more 
controlled flow pattern. This improvement is instrumental in minimizing the turbulence and volatile motion 
that could otherwise affect the tanks' stability and structural integrity. 
Fig.\ref{fig: basic-ring-vertical-vorticity} exhibits the vorticity distribution during a series of sloshing 
events after water impacts the left head of the tanks. When the wave impacting, breaking and rolling occur 
in the tank without baffle, the vorticity in the corresponding fluid regions experiences rapid growth, 
producing apparent local maxima. However, upon the implementation of baffles, the distinctiveness of these 
typical phenomena diminishes, alongside a reduction in vorticity levels. The adoption of baffles in tanks 
appears to mitigate the prominence of these phenomena, effectively suppressing the rapid increase in 
vorticity observed in the absence of baffles.

\begin{figure}
	\centering
	\begin{subfigure}[b]{\textwidth}
		\centering
		\includegraphics[width=4in]{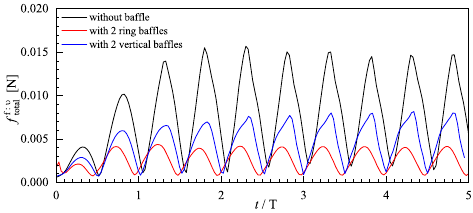}
		\caption{}
		\label{fig: viscous-total}
	\end{subfigure}
        \begin{subfigure}[b]{\textwidth}
		\centering
		\includegraphics[width=4in]{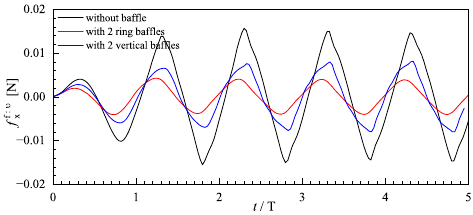}
		\caption{}
		\label{fig: viscous-x}
	\end{subfigure}
        \begin{subfigure}[b]{\textwidth}
		\centering
		\includegraphics[width=4in]{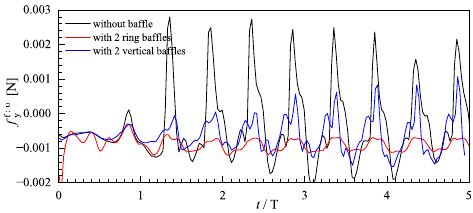}
		\caption{}
		\label{fig: viscous-y}
	\end{subfigure}
	\caption{Comparison of viscous force on tank among normal tank, tank equipped with 2 ring baffles 
 and tank equipped with 2 vertical baffles.}
	\label{fig: viscous-force}
\end{figure}

\begin{figure}
	\centering
	\begin{subfigure}[b]{\textwidth}
		\centering
		\includegraphics[width=4in]{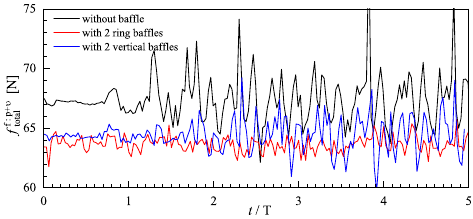}
		\caption{}
		\label{fig: total-total}
	\end{subfigure}
        \begin{subfigure}[b]{\textwidth}
		\centering
		\includegraphics[width=4in]{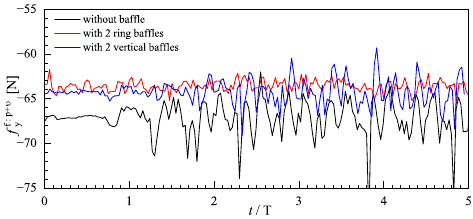}
		\caption{}
		\label{fig: total-y}
	\end{subfigure}
	\caption{Comparison of total force on tank among normal tank, tank equipped with 2 ring baffles 
 and tank equipped with 2 vertical baffles.}
	\label{fig: total-force}
\end{figure}

The viscous force and total force exerted on tank walls are depicted in Fig.\ref{fig: viscous-force} and 
Fig.\ref{fig: total-force}. More intense waves are gradually generated and the forces on the solid walls 
exhibit periodic behavior after the initial period. The viscous force is a variable that intuitively reveals 
the flow state. The viscous component in the x direction accounts for the majority, since the horizontal 
sloshing mainly leads to the x-direction velocity of the fluid. The fluid in the tank without baffle is 
sloshing more violently causing a greater horizontal velocity and viscous force. Small viscous force 
$f^{f:\nu}_{\rm y}$ initially indicates that there is negligible viscous friction caused by the flow in the 
y direction, suggesting the absence of swirl formation at the beginning. Following the simple sway motion, 
the emergence of breaking waves and swirling motion leads to a significant increase in $f^{f:\nu}_{\rm y}$. 
As shown in Fig.\ref{fig: total-force}, the total force in the y direction plays a major role due to gravity. 
A distinct tendency emerges where baffles contribute to a reduction in the total load exerted on the tanks.

\begin{figure}
  \centering
  {\includegraphics[width=5in]{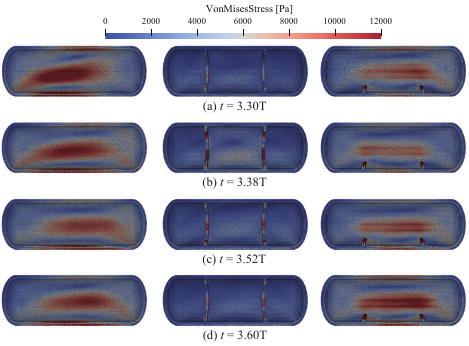}}
  \caption{\label{fig: basic-ring-vertical-stress}{Stress distributions of tank in the third period. 
  left: elastic tank without baffle, middle: elastic tank with 2 ring baffles, right: elastic tank 
  with 2 vertical baffles.}}
\end{figure}

\begin{figure}
  \centering
  {\includegraphics[width=5in]{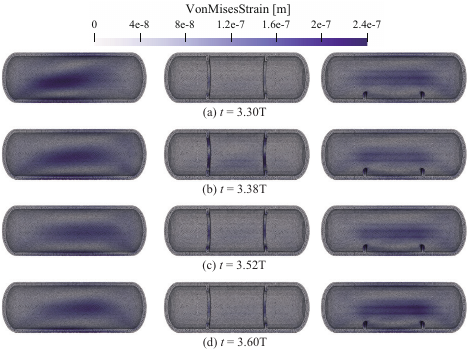}}
  \caption{\label{fig: basic-ring-vertical-strain}{Strain distributions of tank in the third period. 
  left: elastic tank without baffle, middle: elastic tank with 2 ring baffles, right: elastic tank 
  with 2 vertical baffles.}}
\end{figure}

Illustrated in Fig.\ref{fig: basic-ring-vertical-stress} and Fig.\ref{fig: basic-ring-vertical-strain} are 
the distributions of the corresponding stress and strain distributions of tanks in the third period 
(sloshing stable stage). The central regions of both the front and rear walls, along with the bottoms of 
the standard LNG tank experience the highest levels of stress and distortion. As the fluid moves to the 
right and hits the head, a stress concentration zone appears in the lower right corner of the tank. 
These particular zones bear the brunt of the forces from the fluid under operating conditions. 
However, the installation of baffles into the tank design introduces a pivotal shift in how these stresses 
are distributed. These internal baffles alter the flow patterns of the fluid, and play a crucial role in 
alleviating the direct load on the main tank bodies by effectively deflecting the stress concentration domain 
onto themselves. The noteworthy decrease in stress and strain levels within the main bodies demonstrates 
the effectiveness of the baffle system in providing a more even distribution of forces throughout the tank 
interiors. Consequently, the tanks are better equipped to withstand the fluid impact without incurring undue 
wear and tear on the main structure.

To facilitate a more detailed analysis of tank forces, the standard tank at $t$ = 3.38T is taken as 
an example to deconstruct the stress tensor matrix into its components: three normal stresses and 
three shear stresses. At this juncture, the fluid impacts the tank's right head, engendering a 
high-pressure zone in the lower right corner. Concurrently, the stress concentration on the tank 
walls migrates rightward, affected by the fluid's motion. The imposition of displacement constraints 
at the mass center of the tank and the influence of gravitational forces engender pronounced tensile 
stress along the vertical (y-axis) direction, and the horizontal (x-axis) direction, as well as the 
sectional (z-axis) direction, experience significant tensile stress across the tank's upper and lower 
external surfaces. Structural considerations further dictate that the normal stresses within the upper 
and lower internal walls of the z-axis direction exhibit marked compressive stress. Moreover, 
the velocity component of fluid sloshing introduces a shear stress gradient within the x-y plane, 
resulting in structural torsion.

\begin{figure}
  \centering
  {\includegraphics[width=5in]{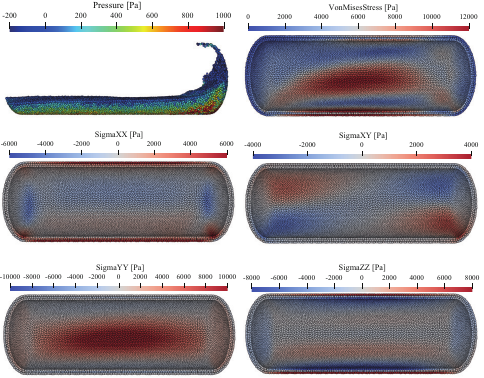}}
  \caption{\label{fig: basic-sigma}{The stress distribution of elastic tank without baffle 
  at $t$ = 3.38T.}}
\end{figure}

\section{Comparison between effects of rigid and elastic baffles} \label{section: rigid-elastic-baffle}

To compare the mitigation effect of rigid and elastic baffles on liquid sloshing, 
water depth $h_w/D = 0.5$ and a thicker wall with 0.03 m are used for modelling in this section, and a vertical baffle is inserted in 
the middle position. A rigid baffle or an elastic baffle is installed in the tank. 
The peripheral portion of the elastic baffle is fixed to the tank wall, and the Young's modulus of the 
elastic baffle is 500 kPa, 50 kPa or 5 kPa respectively. The physical properties of tank material remain 
the same as in the previous section, while the density and Poisson of elastic baffles 
are 2500 kg/m$^3$ and 0.47.

\begin{figure}
  \centering
  {\includegraphics[width=4in]{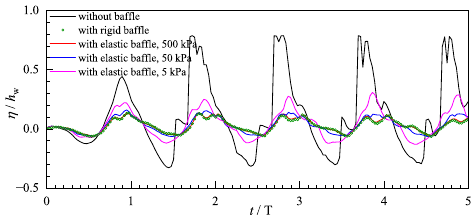}}
  \caption{\label{fig: one-baffle-height}{Comparison of non-dimensional free surface elevation of
  water among normal tank and tank with a rigid or elastic baffle.}}
\end{figure}

\begin{figure}
  \centering
  {\includegraphics[width=5in]{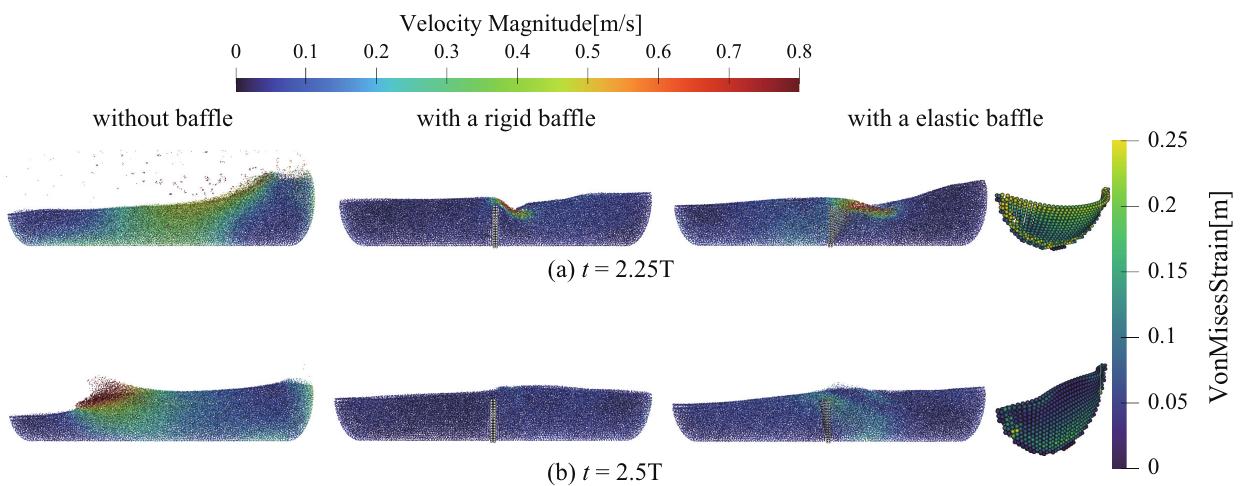}}
  \caption{\label{fig: one-baffle-velocity}{Fluid regimes of water in normal tank and tank with a 
  rigid or elastic baffle (Young's modulus is 5 kPa) at $t$ = 2.25T and $t$ = 2.5T. The right part is the strain distribution of 
  the elastic baffle.}}
\end{figure}

\begin{figure}
  \centering
  {\includegraphics[width=5in]{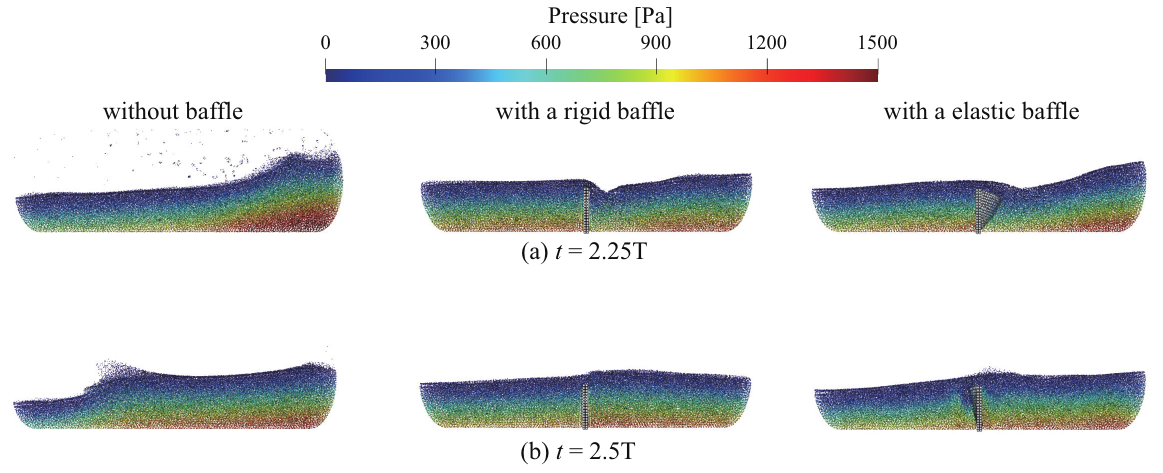}}
  \caption{\label{fig: one-baffle-pressure}{Pressure distribution of water in normal tank and tank
  with a rigid or elastic baffle at $t$ = 2.25T and $t$ = 2.5T.}}
\end{figure}

Fig.\ref{fig: one-baffle-height} shows the variation of free surface height during the first five 
periods, highlighting the critical role of baffles in suppressing large oscillations. However,
when the Young's modulus of the elastic baffle is low, its effectiveness is limited. In other words, 
the closer the baffle's properties are to that of a rigid body, the stronger the inhibitory effect on 
liquid sloshing. As the Young's modulus reaches 500 kPa, there is a negligible difference in the 
performance between elastic and rigid baffles. The corresponding flow patterns of water including 
velocity and pressure distributions in various configurations at 2.25T and 2.5T are shown in 
Fig.\ref{fig: one-baffle-velocity} and Fig.\ref{fig: one-baffle-pressure}, representing the fluid 
hitting the right head and the backflow. The velocity of the free surface is quite high in the tank 
without a baffle, while baffles substantially reduce the velocity magnitude, and have a certain effect 
on improving the high-pressure region of fluid near the tank corner. Also, it can be seen that the 
elastic baffle with Young's modulus equal to 5 kPa deforms under the impact of water, weakening its effectiveness in flow control.

\begin{figure}
  \centering
  \includegraphics[width=4in]{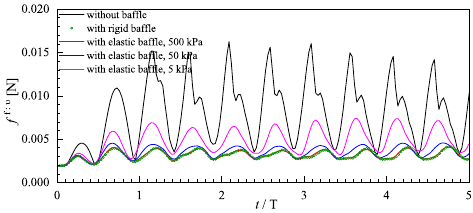}
  \caption{\label{fig: one-baffle-viscous-tank}{Viscous force on tank walls in normal tank and tank 
  with a rigid or elastic baffle.}}
\end{figure}

\begin{figure}
  \centering
  \includegraphics[width=4in]{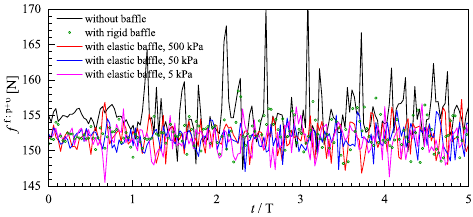}
  \caption{\label{fig: one-baffle-total-tank}{Total force on tank walls in normal tank and tank 
  with a rigid or elastic baffle.}}
\end{figure}

\begin{figure}
  \centering
  \includegraphics[width=5in]{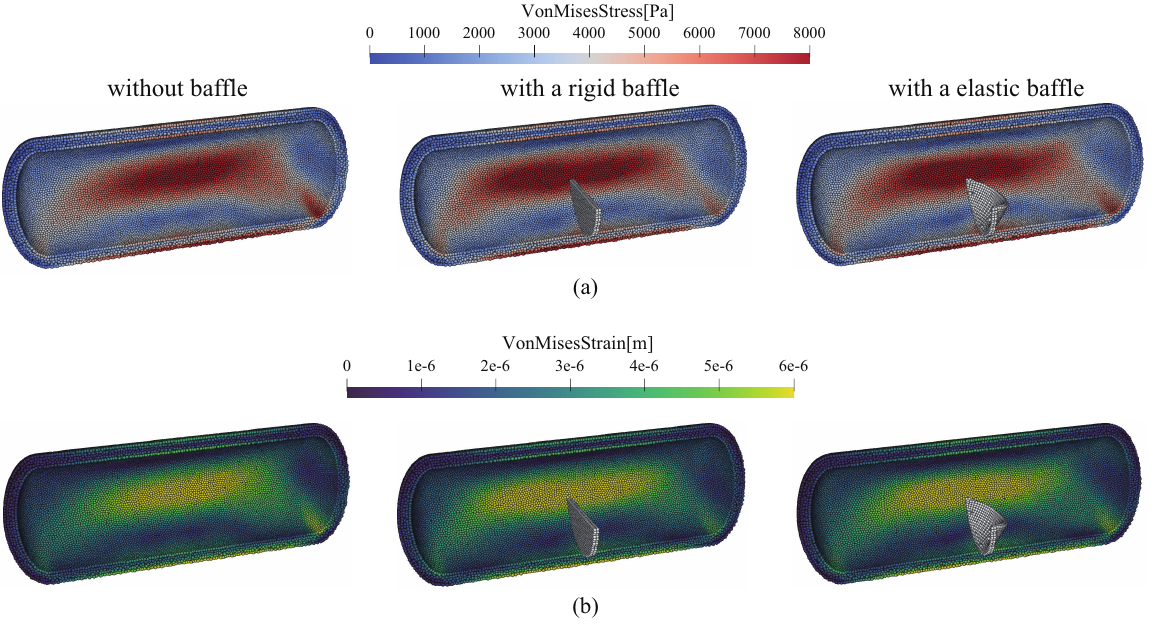}
  \caption{\label{fig: one-baffle-stress-strain-2.25}{Stress and strain distributions of tanks 
  at $t$ = 2.25T.}}
\end{figure}

\begin{figure}
  \centering
  \includegraphics[width=5in]{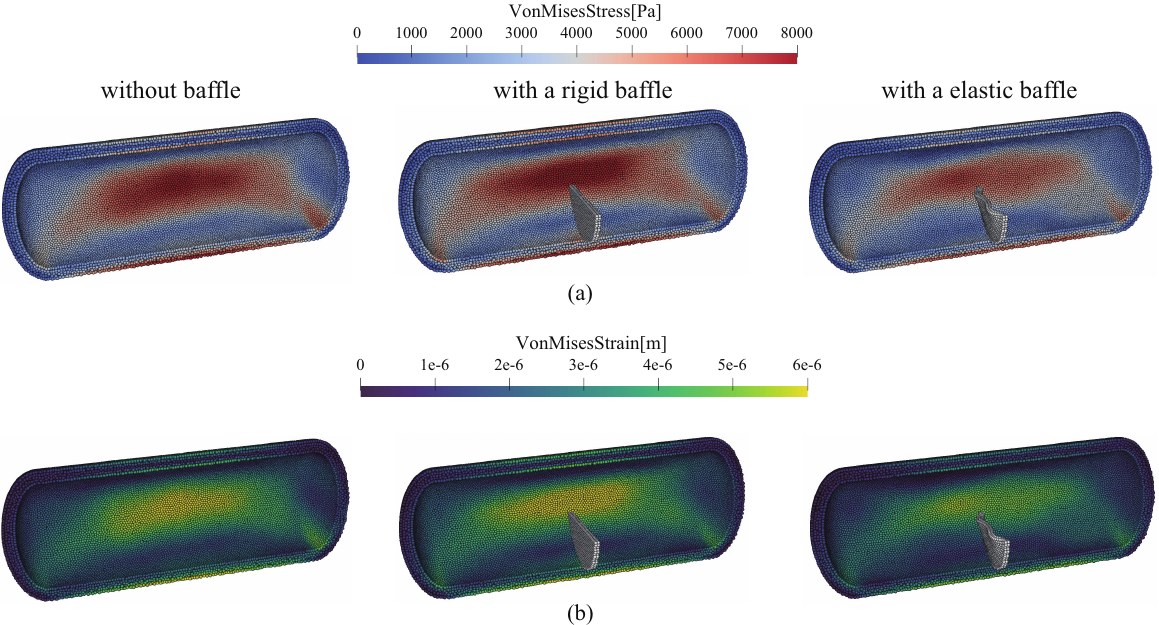}
  \caption{\label{fig: one-baffle-stress-strain-2.5}{Stress and strain distributions of tanks 
  at $t$ = 2.5T.}}
\end{figure}

Fig.\ref{fig: one-baffle-viscous-tank} and Fig.\ref{fig: one-baffle-total-tank}, depicting the viscous 
force and total force of the fluid on tanks, clearly illustrate that baffles have a remarkable effect 
on reducing the viscous force and total force on tanks. In addition, the distributions of stress and 
strain of tank walls at $t$ = 2.25T and $t$ = 2.5T are displayed in 
Fig.\ref{fig: one-baffle-stress-strain-2.25} and Fig.\ref{fig: one-baffle-stress-strain-2.5}. 
The application of baffles results in a reduction in the maximum force and deformation range 
experienced by the tanks to a certain extent.

\begin{figure}
  \centering
  {\includegraphics[width=4in]{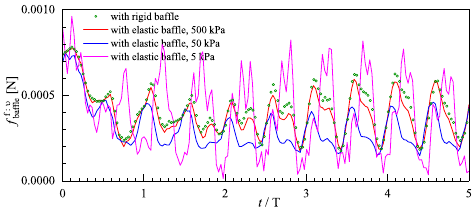}}
  \caption{\label{fig: viscous-force-on-baffle}{Viscous force on the baffle in tank with a rigid or elastic baffle.}}
\end{figure}

\begin{figure}
  \centering
  {\includegraphics[width=4in]{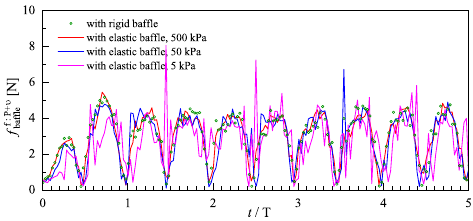}}
  \caption{\label{fig: total-force-on-baffle}{Total force on the baffle in tank with a rigid or elastic baffle.}}
\end{figure}

In the cases configured with various baffles, the viscous force and total force of fluid on the 
baffles are displayed in Fig.\ref{fig: viscous-force-on-baffle} and Fig.\ref{fig: total-force-on-baffle}. 
The baffle with a smaller Young modulus exhibits a more pronounced variation in force over the course of 
the sloshing motion.

\section{Conclusions} \label{section: conclusions}

In the process of LNG transportation, attention should be paid to the load and deformation of the 
tanks with liquid sloshing to avoid structural damage. As a Lagrangian method, the numerical 
research tool based on the SPH method employed in this paper serves as a reliable approach for 
handling the moving interface and free surface. It is necessary to model the tanks with elastic walls 
rather than rigid bodies, and the calculation of the interaction between fluid and solid within a 
single framework is adopted to reduce the error of data transfer and interpolation. In addition, 
the use of two-phase and three-dimensional modelling enhances the accuracy of the simulation. 
The performance of the current model is evaluated through the comparison between experimental data 
and numerical results, demonstrating its reliability. Consequently, this framework is adopted for 
the cases presented in this work to calculate the fluid dynamics for the flow fields and solid 
mechanics for the deformable structure.

The ring baffles and vertical baffles are applied to investigate their anti-sloshing effects and the 
variation of the force exerted on the tank walls by liquid sloshing. The introduction of baffles 
facilitates mitigating sloshing effectively by disrupting wave propagation, and in this comparison case, 
the ring baffles have a more pronounced effect in mitigating large oscillations. Examining the stress 
and strain distributions within the tanks reveals that due to the existence of baffles, concentration 
regions are shifted from the main tank bodies to the baffles, thereby reducing the risk of structural 
damage to the primary tank structure.

Finally, the rigid baffle and elastic baffles with several Young's moduli are inserted in the middle 
of the tank to investigate the influence of baffle material properties on sloshing inhibition. 
Numerical results indicate that baffle properties resembling those of a rigid body are more effective 
in suppressing liquid sloshing, resulting in a more noticeable improvement in the forces acting on the 
tank walls. This is because an elastic baffle with a high Young's modulus exhibits weaker control 
over fluid flow.

%
%
\section*{Acknowledgement}
C.X. Zhao is fully supported by the China Scholarship Council (CSC) (No:202206280028).
Y.C. Yu is fully supported by the China Scholarship Council (CSC) (No:201806120023).
X.Y. Hu would like to express their gratitude to Deutsche Forschungsgemeinschaft for their sponsorship 
of this research under grant number DFG HU1572/10-1 and DFG HU1527/12-1. 
%
%
\section*{CRediT authorship contribution statement}
{\bfseries  Chenxi Zhao:} Investigation, Methodology, Visualization, Validation, Formal analysis, Writing - 
original draft, Writing - review \& editing;
{\bfseries Yan Wu:} Investigation, Methodology, Validation;
{\bfseries  Yongchuan Yu:} Investigation, Methodology; 
{\bfseries  Xiangyu Hu:} Supervision, Methodology, Writing - review \& editing;
{\bfseries  Oskar J. Haidn:} Supervision.
%
%
\section*{Declaration of competing interest }
The authors declare that they have no known competing financial interests 
or personal relationships that could have appeared to influence the work reported in this paper.
%
%
\section*{Data availability}
The code is open source on \href{https://www.sphinxsys.org}{https://www.sphinxsys.org}.
%
%
\clearpage
\bibliography{reference}
%
%
\end{document}